\documentclass{article}
\usepackage{epsfig}
\usepackage{axodraw}
\def\simlt{\stackrel{<}{{}_\sim}}
\def\simgt{\stackrel{>}{{}_\sim}}

\renewcommand{\baselinestretch}{1.2}

\begin{document}

\thispagestyle{empty}

{\normalsize\sf
\rightline {hep-ph/0207242}
\rightline{TUM-HEP-472/02}
\rightline{IFT-02/26}
\vskip 3mm
\rm\rightline{July 2002}
}

\vskip 5mm

\begin{center}
{\LARGE\bf Supersymmetry (at large $\tan\beta$) and flavour 
physics}\footnote{Talk given by P.H.Ch. at the International Meeting
{\it Physics from the Planck Scale to the Electroweak Scale}, May 2002,
Kazimierz Dolny, Poland and dedicated to Stefan Pokorski on his 60$^{th}$ 
birthday. To apper in the special volume of {\sl Acta Physica Polonica} 
{\bf B}.}

\vskip 10mm

{\large\bf Piotr H.~Chankowski$^1$ and Janusz Rosiek$^{1,2}$} \\[5mm]

{\small $^1$ Institute of Theoretical Physics, Warsaw University}\\
{\small Ho\.za 69, 00-681 Warsaw, Poland}
{\small $^2$ Physik Department, Technische Universit{\"a}t M{\"u}nchen,}\\
{\small D-85748 Garching, Germany}\\

\end{center}

\vskip 5mm

\renewcommand{\baselinestretch}{1.1}

\begin{abstract}
  Recent development in exploring flavour dynamics in the
  supersymmetric extension of the Standard Model is reviewed. Emphasis
  is put on possible interesting effects in $b$-physics arising for
  large values of $\tan\beta$ both in the case of minimal flavour
  violation and in the case of flavour violation originating in the
  sfermion sector. The importance of the flavour changing neutral
  Higgs boson couplings generated by the scalar penguin diagrams and
  their role in the interplay of neutral $B$-meson mixing and
  $B^0_{d,s}\rightarrow\mu^+\mu^-$ decays is discussed. It is pointed
  out that observation of the $B^0_d\rightarrow\mu^+\mu^-$ decay with
  BR at the level $\simgt3\times10^{-8}$ would be a strong indication of
  nonminimal flavour violation in the quark sector. Possible impact of
  flavour violation in the slepton sector on neutrino physics is also
  discussed.
\end{abstract}
\renewcommand{\baselinestretch}{1.2}

\newpage 
\setcounter{page}{1} 
\setcounter{footnote}{0}
\section{Introduction}
\setcounter{equation}{0} 

Physics of flavour and of CP violation continues to be an interesting
subject to study in various extensions of the Standard Model (SM). On
one hand, before the advent of LHC and linear colliders, which will
enable us to probe energies much above the electroweak scale directly,
rare processes intensively studied in numerous experiments are the
first place where the effects of new physics - i.e. virtual effects of
new particles - can most likely be detected. On the other hand,
studies of baryogenesis \cite{BARIOG} strongly suggest that the SM
with its unique source of CP violation and known particle content is
unable to explain the baryon to photon density number ratio,
$n_B/n_\gamma\sim10^{-9}$, observed in the Universe, given the present
lower limit on the Higgs boson mass. This supports expectations that
some deviations from the SM predictions will eventually be encountered
in the ongoing or planned precision studies of rare and CP violating
processes. Finally, the first direct indication of inadequacy of the
SM has also to do with flavour physics - namely with neutrino
oscillations. Explanation of the observed neutrino oscillations
requires the introduction of flavour mixing (and perhaps also of CP
violation) in the lepton sector. While the SM can easily be extended
to describe neutrino oscillations, there are strong theoretical
arguments that the observed phenomena have their origin in physics at
energy scales much higher than the electroweak scale.

Physics of flavour in the framework of supersymmetric extension of the
Standard Model was also always in the center of Stefan Pokorski's interest. 
It is therefore a pleasure to devote this article to him. The subject is 
of course too vast to be reviewed here in all details. Instead we 
concentrate on its most interesting in our opinion aspects. These include 
recent investigations of the supersymmetric effects in $b$-physics arising
for large $\tan\beta$ and in the neutrino sector. In this context it is 
appropriate to recall here that systematic investigations of supersymmetry at 
large $\tan\beta$ begun with the Stefan Pokorski's seminal paper \cite{OLPO}.

\section{Flavour violation: minimal and generalized minimal }

Extensions of the SM can be divided into two broad classes: models in
which the Cabibbo-Kobayashi-Maskawa matrix (CKM) in the quark sector
and Maki-Nakagawa-Sakata matrix (MNS) in the lepton sector are the
only sources of flavour and CP violation and models in which there are
entirely new sources of flavour and/or CP violation. Both options can
be realized independently in the quark and lepton sectors of the
simplest supersymmetric extension of the SM - the MSSM - and it is the
experimental task of utmost importance to establish which one is
realized in Nature.

If the CKM matrix is the only source of flavour and CP violation in
the quark sector, the natural question is what is the impact of new
physics on the determination of its elements. In particular one wants
to know the value of the $V_{td}$ element which is needed e.g. to
predict the rate of the decay $B^0_d\rightarrow\mu^+\mu^-$ and other
interesting rare processes. It is also important to see if the
consistency of the determination of the CKM matrix elements from
different processes imposes any constraints on the MSSM parameters.

The CKM matrix $V$ is most conveniently parameterized as follows \cite{WOL}:
\begin{eqnarray}
V= \left(\matrix{1-\lambda^2/2&\lambda & A\lambda^3(\rho - i\eta)\cr
-\lambda & 1-\lambda^2/2 &A\lambda^2 \cr
A\lambda^3(1 - \rho - i\eta)&-A\lambda^2&1}\right)+{\cal O}(\lambda^4)
\nonumber
\end{eqnarray}
The Wolfenstein parameters $\lambda\approx0.222\pm0.0018$ and
$A\approx0.83\pm0.06$ are rather accurately determined from
transitions dominated by tree level contributions, and are hence
insensitive to new physics. At present, processes of this kind put
also some constraints on the remaining two (conveniently rescaled
\cite{BULAOS}) parameters $\bar\rho\equiv\rho(1-\lambda^2/2)$ and
$\bar\eta\equiv\eta(1-\lambda^2/2)$.  The value of the combination
$R_b\equiv\sqrt{\bar\rho^2+\bar\eta^2}$ is constrained to $0.27\simlt
R_b\simlt0.46$ by the result $|V_{ub}|/|V_{cb}|=0.08\pm0.02$ (at 95 \%
C.L.) extracted from the charmless $B$ decays. The CP violating time
dependent asymmetry in the $B\rightarrow\psi K_S$ decay constrains the
phase $\beta_{\rm ut}$ of the $V_{td}$ element: $V_{td} =
|V_{td}|e^{-i\beta_{\rm ut}}$. In minimal models this asymmetry is
simply given by $\sin2\beta_{\rm ut}$. The average of the measurements
done at BaBar and Belle gives $\sin2\beta_{\rm ut}=
2\bar\eta(1-\bar\rho)/\sqrt{(1-\bar\rho)^2+\bar\eta^2}=0.78\pm0.08$
\cite{ALEXAN}. There are also prospects for extracting from such
processes also the phase $\gamma_{\rm ut}$ of the $V_{ub}$ element:
$V_{ub}=|V_{ub}|e^{-i\gamma_{\rm ut}}$. However, large theoretical and
experimental uncertainties still prevent precise determination of
$\bar\rho$ and $\bar\eta$ exclusively from tree level dominated
processes.

Parameters $\bar\rho$ and $\bar\eta$ are also extracted from
measurements of the $B^0_{s,d}$-$\bar B^0_{s,d}$ meson mass
differences $\Delta M_{s,d}$ and of the parameter $\varepsilon_K$ of
CP violation in the neutral kaon system. This allows to overconstrain
the values of $\bar\rho$ and $\bar\eta$ and test the assumption of
minimal flavour and CP violation in the quark sector. However, since
all the three quantities are loop induced, the new physics can
contribute to relevant amplitudes. Therefore, the values of $\bar\rho$
and $\bar\eta$ determined from $\Delta M_{s,d}$ and $\varepsilon_K$
can significantly depend on new physics. One can also expect that
consistency of the CKM parameters determination puts some constraints
on new physics.

To compute $\Delta M_{s,d}$ and $\varepsilon_K$ one integrates out
from the theory all the states with masses $\simgt M_W$ and constructs
the effective Hamiltonian of the form
\begin{eqnarray}
{\cal H}_{\rm eff} = {G^2_FM^2_W\over16\pi^2}
\sum_X \lambda^X_{\rm CKM} C_X{\cal O}_X\label{eqn:heff}
\end{eqnarray}
where ${\cal O}_X$ are the local four-quark operators ($X$ labels
different Lorentz structures: $X=$VLL, VRR, VLR, SLL, SRR, SLR, TL,
and TR) and $\lambda^X_{\rm CKM} C_X$ are their Wilson coefficients.
An important feature of the minimal flavour violation is the
factorization of the Wilson coefficients into $\lambda^X_{\rm CKM}$
which depends only on the CKM matrix elements and $C_X$ which to (a
good approximation) are real numbers.

At the level of the effective Hamiltonian (\ref{eqn:heff}) models of
new physics in which the CKM matrix is the only source of flavour and
CP violation in the quark sector can be further divided into two broad
classes \cite{BUER}:
\begin{itemize}
\item the MFV (minimal flavour violation) models - truly minimal ones,
  in which, just as in the SM, only $C_{\rm VLL}$ Wilson coefficient
  is non-negligible and $C_{\rm VLL}$ responsible for
  $B^0_{s,d}$-$\bar B^0_{s,d}$ and $K^0$-$\bar K^0$ meson mixing are
  all equal (universal value of $C_{\rm VLL}$).
\item the GMFV (generalized minimal flavour violation) models - in
  which more $C_X$ are non-negligible and/or are non-universal.
\end{itemize}
As we shall see, the MSSM can be of either type, depending on the
ratio $v_2/v_1\equiv\tan\beta$ of the vacuum expectation values of the
two Higgs boson doublets.

Basic formulae used to determine $\bar\rho$ and $\bar\eta$ read (see
e.g. refs. \cite{BUREV,BUER} for further details):
\begin{eqnarray}
\bar\eta\left[(1-\bar\rho)A^2\eta_2F^{\varepsilon}+P_c\right]
A^2\hat B_K=0.204~,\label{eqn:epsK}
\end{eqnarray}
where the number on the rhs stems from the measured value 
$\varepsilon_K=2.28\times10^{-3}$, and
\begin{eqnarray}
\Delta M_d={G^2_FM^2_W\over16\pi^2}M_{B_d}\eta_B\hat B_{B_d}F^2_{B_d}
|V_{tb}^\ast V_{td}|^2 |F^d|
\propto\hat B_{B_d}F^2_{B_d} |(1-\bar\rho)-i\bar\eta|^2 |F^d|
\nonumber\\
\Delta M_s={G^2_FM^2_W\over16\pi^2}M_{B_s}\eta_B\hat B_{B_s}F^2_{B_s}
|V_{tb}^\ast V_{ts}|^2 |F^s|
\propto\hat B_{B_s}F^2_{B_s}  F^s ~.\phantom{aaaaa}
\label{eqn:dMB}
\end{eqnarray}
In eqs. (\ref{eqn:epsK}), (\ref{eqn:dMB}) $\eta_2=0.57$ and
$\eta_B=0.55$ summarize the short distance QCD corrections to $C_{\rm
  VLL}$ Wilson coefficients and $P_c=0.30\pm0.05$ is the known charmed
quark loop contribution to $\varepsilon_K$. Factors $\hat
B_K\approx0.85\pm0.15$, $\hat B_{B_d}F^2_{B_d}\approx(230\pm40{\rm
  ~MeV})^2$ and $\hat B_{B_s}F^2_{B_s}\approx(265\pm40{\rm ~MeV})^2$
\cite{LATT}
parameterizing matrix elements of the standard VLL operators are the
biggest sources of uncertainties. The three factors $F^{\varepsilon}$,
$F^d$ and $F^s$ can be expressed in terms of the Wilson coefficients
$C_X$, their QCD RG running and matrix elements of the operators
${\cal O}_X$ for $X\neq$~VLL. In a concrete model of new physics such
as e.g. the MSSM, $F^{\varepsilon}$, $F^d$ and $F^s$ are calculable
functions of its parameters.\footnote{All the matrix elements of the
  operators ${\cal O}_X$, also for $X\neq$VLL, are now known from
  lattice calculations \cite{AL,MARTETAL}. Nevertheless, the
  uncertainties in their values still introduce some uncertainty in
  the factors $F^i$ which depend, apart from Wilson coefficients and
  calculable QCD RG factors, also on the ratios of these matrix
  elements to the matrix element of the standard VLL operator.}  The
distinction between MFV and GMFV models is reflected in that in the
formers $F^{\varepsilon}=F^d=F^s$ whereas in the latter models all
$F^i$ can be different. In the SM $F^{\varepsilon}=F^d=F^s= F_{\rm
  SM}=S_0(\bar m_t)\approx 2.38\pm0.11$ for $\bar m_t(m_t)=166\pm5$
GeV.

The measured $B^0_d$-$\bar B^0_d$ mass difference, $\Delta
M_d=0.496/$ps, puts the constraint\footnote{All bounds and allowed
  ranges of various quantities quoted in this article are obtained by
  scanning over all uncertainties within their respective 1$\sigma$
  ranges.}  $1.04\simlt\sqrt{|F^d|}R_t\equiv\sqrt{|F^d|}|1-\bar\rho -
i\bar\eta|\simlt1.69$.  The role of $\Delta M_s$ which does not depend
directly on $\bar\rho$ and $\bar\eta$ is twofold. Firstly, as follows
from eqs. (\ref{eqn:dMB}), any new physics model must be such that
$F^s$ it gives rise to satisfies \cite{BUCHROSL}
\begin{eqnarray}
0.52 \left({\Delta M_s\over15/{\rm ps}}\right)<\left|{F^s\over F_{\rm SM}}
\right|< 1.29 \left({\Delta M_s\over15/{\rm ps}}\right)\label{eqn:fslim}
\end{eqnarray}
Since at present only the lower limit on the $B^0_s$-$\bar B^0_s$ mass
difference is known, $\Delta M_s>15/{\rm ps}$, the factor $F^s$ is
bounded only from below. Secondly, once measured, $\Delta M_s$
combined with $\Delta M_d$ will allow for more precise determination
of $|V_{td}|\propto|1-\bar\rho - i\bar\eta|$ because the ratio $\xi^2$
of $F^2_{B_s}\hat B_{B_s}$ to $F^2_{B_d}\hat B_{B_d}$ is known with
better accuracy than these factors individually: $\xi=1.15\pm0.06$
\cite{FLSACH}.  For given $F^s/F^d$, the value of $R_t$ is then
determined, from the formula
\begin{eqnarray}
R_t\equiv|1-\bar\rho - i\bar\eta|=
0.82 ~\xi\left({15/{\rm ps}\over\Delta M_s}\right)
\sqrt{\left|{F^s\over F^d}\right|}\nonumber
\end{eqnarray}
Note that $R_t$ determined in this way is universal in the whole class
of MFV models for which $F^s/F^d=1$. In contrast, in GMFV models the
extracted value of $R_t$ does depend on new physics contributions to
$F^s$ and/or $F^d$.

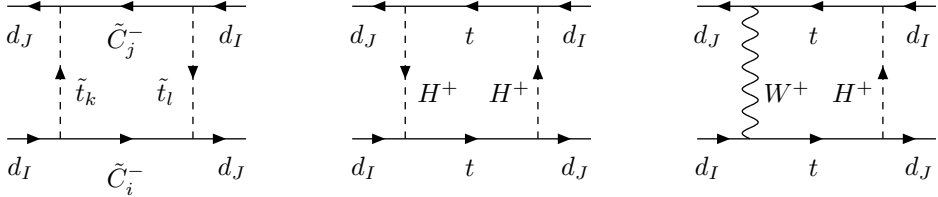
\begin{figure}[htbp]
\begin{center}
\begin{picture}(350,80)(0,0)
\ArrowLine(0,20)(20,20)
\ArrowLine(20,20)(70,20)
\ArrowLine(70,20)(90,20)
\ArrowLine(90,70)(70,70)
\ArrowLine(70,70)(20,70)
\ArrowLine(20,70)(0,70)
\DashArrowLine(20,20)(20,70){3}
\DashArrowLine(70,70)(70,20){3}
\Text(5,9)[]{$d_I$}
\Text(85,9)[]{$d_J$}
\Text(5,59)[]{$d_J$}
\Text(85,59)[]{$d_I$}
\Text(45,5)[]{$\tilde C^-_i$}
\Text(45,57)[]{$\tilde C^-_j$}
\Text(30,38)[]{$\tilde t_k$}
\Text(60,38)[]{$\tilde t_l$}
\ArrowLine(130,20)(150,20)
\ArrowLine(150,20)(200,20)
\ArrowLine(200,20)(220,20)
\ArrowLine(220,70)(200,70)
\ArrowLine(200,70)(150,70)
\ArrowLine(150,70)(130,70)
\DashArrowLine(150,70)(150,20){3}
\DashArrowLine(200,20)(200,70){3}
\Text(135,9)[]{$d_I$}
\Text(215,9)[]{$d_J$}
\Text(135,59)[]{$d_J$}
\Text(215,59)[]{$d_I$}
\Text(175,9)[]{$t$}
\Text(175,59)[]{$t$}
\Text(163,38)[]{$H^+$}
\Text(190,38)[]{$H^+$}
\ArrowLine(260,20)(280,20)
\ArrowLine(280,20)(330,20)
\ArrowLine(330,20)(350,20)
\ArrowLine(350,70)(330,70)
\ArrowLine(330,70)(280,70)
\ArrowLine(280,70)(260,70)
\Photon(280,20)(280,70){3}{6}
\DashArrowLine(330,20)(330,70){3}
\Text(265,9)[]{$d_I$}
\Text(345,9)[]{$d_J$}
\Text(265,59)[]{$d_J$}
\Text(345,59)[]{$d_I$}
\Text(305,9)[]{$t$}
\Text(305,59)[]{$t$}
\Text(295,38)[]{$W^+$}
\Text(320,38)[]{$H^+$}
\end{picture}
\end{center}
\caption{Contribution of the chargino-stop and charged Higgs ($W^\pm$) 
  boson box diagrams to $F^\varepsilon$, $F^d$ and $F^s$ in the MSSM.
  Crossed diagrams are not shown.}
\label{fig:boxes}
\end{figure}

\section{Supersymmetric contributions to $F^\varepsilon$, $F^s$ and $F^d$}

Dominant supersymmetric contributions to $F^\varepsilon$, $F^s$ and
$F^d$ for small and moderate values of the ratio of the two MSSM Higgs
boson doublets vacuum expectation values $v_2/v_1\equiv\tan\beta$ are
well studied \cite{BRFEZW,MIPORO}. They arise from box diagrams shown
in fig.  \ref{fig:boxes} and for $2\simlt\tan\beta\simlt20$ give
$F^{\varepsilon}=F^d=F^s\equiv F$. Thus, for not too large values of
$\tan\beta$ the MSSM is of the MFV type.  Maximal values $F/F_{\rm
  SM}\simgt1.4$ are reached for lightest sparticles still not excluded
by direct supersymmetry searches and $\tan\beta$ as small as
possible.\footnote{Recall also, that for the top squarks lighter than
  $\simlt1$ TeV the range $1\simlt\tan\beta\simlt2$ is excluded by the
  unsuccessful search of the lightest Higgs boson at LEP.}  With
increasing $\tan\beta$ and/or increasing sparticle and charged Higgs
boson masses the value of $F/F_{\rm SM}$ decreases to 1
\cite{BRFEZW,MIPORO}.

\begin{figure}[htbp]
\begin{center}
\begin{picture}(340,80)(0,0)
\ArrowLine(10,10)(50,20)
\ArrowLine(50,20)(90,10)
\Vertex(50,20){5}
\ArrowLine(50,60)(10,70)
\ArrowLine(90,70)(50,60)
\Vertex(50,60){5}
\DashLine(50,20)(50,60){3}
\Text(78,40)[]{\small $h^0$,$H^0$,$A^0$}
\Text(25,0)[]{\small $(d_L)_I$}
\Text(75,0)[]{\small $(d_R)_J$}
\Text(75,75)[]{\small $(d_L)_I$}
\Text(25,75)[]{\small $(d_R)_J$}
\ArrowLine(130,10)(170,20)
\ArrowLine(170,20)(210,10)
\Vertex(170,20){5}
\ArrowLine(170,60)(130,70)
\ArrowLine(210,70)(170,60)
\Vertex(170,60){5}
\DashLine(170,20)(170,60){3}
\Text(198,40)[]{$h^0$,$H^0$,$A^0$}
\Text(145,0)[]{$(d_R)_I$}
\Text(195,0)[]{$(d_L)_J$}
\Text(195,75)[]{$(d_R)_I$}
\Text(145,75)[]{$(d_L)_J$}
\ArrowLine(250,10)(290,20)
\ArrowLine(290,20)(330,10)
\Vertex(290,20){5}
\ArrowLine(290,60)(250,70)
\ArrowLine(330,70)(290,60)
\Vertex(290,60){5}
\DashLine(290,20)(290,60){3}
\Text(318,40)[]{$h^0$,$H^0$,$A^0$}
\Text(265,0)[]{$(d_L)_I$}
\Text(315,0)[]{$(d_R)_J$}
\Text(315,75)[]{$(d_R)_I$}
\Text(265,75)[]{$(d_L)_J$}
\end{picture}
\end{center}
\caption{Double penguin diagram contributing to $C_1^{\rm SLL}$,
$C_1^{\rm SRR}$ and $C_2^{\rm LR}$ Wilson coefficients, respectively 
in the MSSM with large $\tan\bar\beta$.}
\label{fig:2pg}
\end{figure}
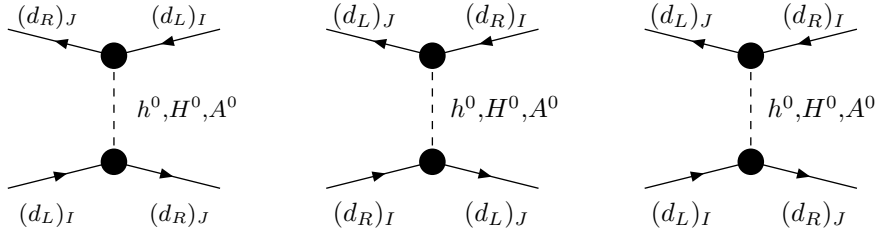

\begin{figure}[htbp]
\begin{center}
\begin{picture}(350,85)(0,0)
\ArrowLine(0,20)(20,20)
\ArrowLine(45,20)(20,20)
\ArrowLine(45,20)(70,20)
\ArrowLine(90,20)(70,20)
\DashArrowLine(45,60)(20,20){3}
\DashArrowLine(45,60)(70,20){3}
\DashArrowLine(45,60)(45,80){3}
\Text(57,70)[]{\small $H^u$}
\Text(5,9)[]{\small $d^c_J$}
\Text(85,9)[]{\small $q_I$}
\Text(30,9)[]{\small $\tilde H^d$}
\Text(55,9)[]{\small $\tilde H^u$}
\Text(20,40)[]{\small $\tilde q$}
\Text(70,40)[]{\small $\tilde u^c$}
\Text(45,-5)[]{\bf a)}
\Line(43,18)(47,22)
\Line(43,22)(47,18)
\ArrowLine(130,20)(150,20)
\ArrowLine(175,20)(150,20)
\ArrowLine(175,20)(200,20)
\ArrowLine(220,20)(200,20)
\DashArrowLine(150,20)(162,40){3}
\DashArrowLine(162,40)(175,60){3}
\DashArrowLine(200,20)(175,60){3}
\DashArrowLine(175,60)(175,80){3}
\Line(173,18)(177,22)
\Line(173,22)(177,18)
\Line(160,38)(164,42)
\Line(160,42)(164,38)
\Text(187,70)[]{\small $H^u$}
\Text(135,9)[]{\small $d^c_J$}
\Text(215,9)[]{\small $q_I$}
\Text(160,9)[]{$\tilde g$}
\Text(185,9)[]{$\tilde g$}
\Text(150,35)[]{\small $\tilde d^c$}
\Text(160,50)[]{\small $\tilde d^c$}
\Text(200,40)[]{\small $\tilde q$}
\Text(175,-5)[]{\bf b)}
\ArrowLine(260,20)(280,20)
\ArrowLine(305,20)(280,20)
\ArrowLine(305,20)(330,20)
\ArrowLine(350,20)(330,20)
\DashArrowLine(280,20)(305,60){3}
\DashArrowLine(330,20)(318,40){3}
\DashArrowLine(318,40)(305,60){3}
\DashArrowLine(305,60)(305,80){3}
\Line(303,18)(307,22)
\Line(303,22)(307,18)
\Line(316,38)(320,42)
\Line(316,42)(320,38)
\Text(317,70)[]{\small $H^u$}
\Text(265,9)[]{\small $d^c_J$}
\Text(345,9)[]{\small $q_I$}
\Text(290,9)[]{$\tilde g$}
\Text(315,9)[]{$\tilde g$}
\Text(280,40)[]{\small $\tilde d^c$}
\Text(333,35)[]{\small $\tilde q$}
\Text(323,50)[]{\small $\tilde q$}
\Text(305,-5)[]{\bf c)}
\end{picture}
\end{center}
\caption{Diagrams generating flavour changing neutral Higgs boson couplings.
  $\tilde q$, $\tilde u^c$ and $\tilde d^c$ denote electroweak
  eigenstates.  Diagrams b) and c) contribute only in the case of
  non-minimal flavour violation arising from squark mass matrices
  (sec. 5).}
\label{fig:scpg}
\end{figure}
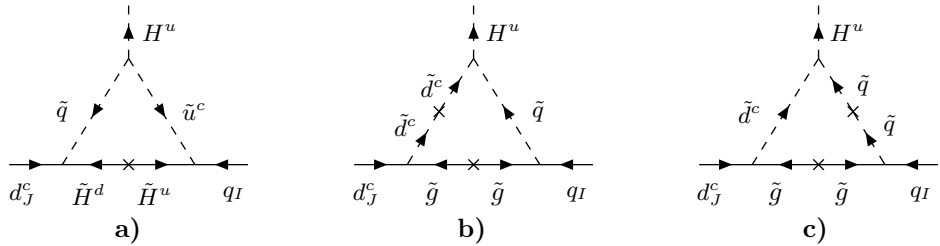

As has been found recently \cite{HAPOTO,BAKO,BUCHROSL} (see also
\cite{DUBNA}), for large values of $\tan\beta$, $\sim50$, the Wilson
coefficient $C_{\rm SLL}$, $C_{\rm SRR}$ and $C_{\rm SLR}$ of the
effective Hamiltonian (\ref{eqn:heff}) can receive very large
contributions from the so-called double scalar penguin diagrams
(formally two-loop) shown in fig. \ref{fig:2pg}. The origin of the
flavour changing couplings of the neutral Higgs bosons can be easiest
understood in the effective Lagrangian approach
\cite{CHPO,HAPOTO,CHSL} (see also \cite{BUCHROSL2}): due to the
triangle (scalar penguin) diagram shown in fig.  \ref{fig:scpg}a,
integrating out sparticles (but not the Higgs bosons) in the
approximation of unbroken electroweak symmetry generates the Yukawa
coupling of the $H^u$ Higgs doublet to down-type quarks that is not
present in the original MSSM Lagrangian. Thus, in the low energy
effective Lagrangian both Higgs doublets, $H^d$ and $H^u$, couple to
down-type quarks and this, after the electroweak symmetry breaking,
gives rise to the tree level flavour changing couplings of $A^0$,
$h^0$ and $H^0$.

For the transitions $d_I\bar d_J\leftrightarrow d_J\bar d_I$ the
double penguin diagrams give
\begin{eqnarray}
&&C^{\rm SLL}=-{\alpha_{\rm em}\over4\pi s^2_W}{m^4_t\over M^4_W}
m_{d_J}^2 X_{tC}^2 \tan^4\beta \times\left({\cos^2\alpha\over M^2_H} 
+ {\sin^2\alpha\over M^2_h}- {\sin^2\beta\over M^2_A}\right)
\nonumber\\ 
&&C^{\rm SLR}=-{\alpha_{\rm em}\over2\pi s^2_W}{m^4_t\over M^4_W}
m_{d_J}m_{d_I} X_{tC}^2 \tan^4\beta \times\left({\cos^2\alpha\over M^2_H} 
+ {\sin^2\alpha\over M^2_h}+ {\sin^2\beta\over M^2_A}\right).\nonumber
\label{eqn:babucor}
\end{eqnarray}
($C^{\rm SRR}$ is obtained from $C^{\rm SLL}$ by replacing $m_{d_J}^2$
by $m_{d_I}^2$). $X_{tC}$ is given by
\begin{eqnarray}
X_{tC} = \sum_{j=1}^2Z_+^{2j}Z_-^{2j}{A_t\over m_{C_j}}
H_2(x^{t/C_j}_1,x^{t/C_j}_2),\nonumber
\end{eqnarray}
where $x^{t/C_j}_i=M^2_{\tilde t_i}/m^2_{C_j}$, $i,j=1,2$ are the
ratios of the stop and chargino masses squared, the matrices $Z_+$ and
$Z_-$ are defined in ref.~\cite{ROS} and
\begin{eqnarray}
H_2(x,y)={x\ln x\over(1-x)(x-y)}+{y\ln y\over(1-y)(y-x)}\nonumber 
\end{eqnarray}
Because for $M_A>M_Z$ and $\tan\beta\gg1$ one has $M^2_H\approx
M^2_A$, $\sin\alpha\approx0$, the coefficients $C^{\rm SLL}$ and
$C^{\rm SRR}$ are suppressed \cite{BAKO}. It turns out however, that
for sufficiently large stop mixing parameter $A_t$ the double penguin
contribution to $C^{\rm SLR}$ for the $b\bar s\leftrightarrow s\bar b$
transition is significant despite the suppression by the strange quark
mass. Inserting numbers one finds
\begin{eqnarray}
C^{\rm SLR}\approx-4.64\times\left({200 ~{\rm GeV}\over M_A}\right)^2
\left({\tan\beta\over50}\right)^4X^2_{tC}\nonumber
\end{eqnarray}
for $m_b=2.7$ GeV, $m_s=60$ MeV at the scale $Q=m_t$, $M_H=M_A$ and
$\sin\alpha=0$. For $\tan\beta\sim50$, $X_{tC}\sim{\cal O}(1)$ and
CP-odd Higgs boson not too heavy this is comparable with the value of
the Wilson coefficient of the standard VLL operator: $C^{\rm
  VLL}=4S_0(\bar m_t)\approx9.5$. The ratio $C^{\rm SLR}/C^{\rm VLL}$
is further increased by the QCD RG effects \cite{BUJAUR}: $C^{\rm
  SLR}(4.6 ~{\rm GeV})=2.23 ~C^{\rm SLR}(m_t)$ while $C^{\rm VLL}(4.6
~{\rm GeV})=0.84 ~C^{\rm VLL}(m_t)$.  For the transitions $b\bar
d\leftrightarrow d\bar b$ and $d\bar s\leftrightarrow s\bar d$ similar
double penguin contributions to $C^{\rm SLR}$ are negligible being
suppressed by $m_d/m_s\approx0.06$ and $m_d/m_b\approx0.001$,
respectively. Thus, for large values of $\tan\beta$ the MSSM becomes
of the GMFV type with $F^\varepsilon\approx F^d\approx F_{\rm SM}\neq
F^s$ and $F^s/F_{\rm SM}<1$.

The important features of the double penguin contribution to $C^{\rm
  SLR}$ are the following: it grows as $\tan^4\beta$, it is always
negative leading to $F^s< F_{\rm SM}$ and is directly sensitive to the
top squarks mixing ($C^{\rm SLR}\propto A_t$). Moreover it does not
vanish if all the sparticle mass parameters are uniformly scaled up
(non-decoupling effect). It does however vanish as the inverse square
of the Higgs sector mass scale set by $M_A$.

\begin{figure}[htbp]
\begin{center}
\begin{tabular}{p{0.72\linewidth}}
\mbox{\epsfig{file=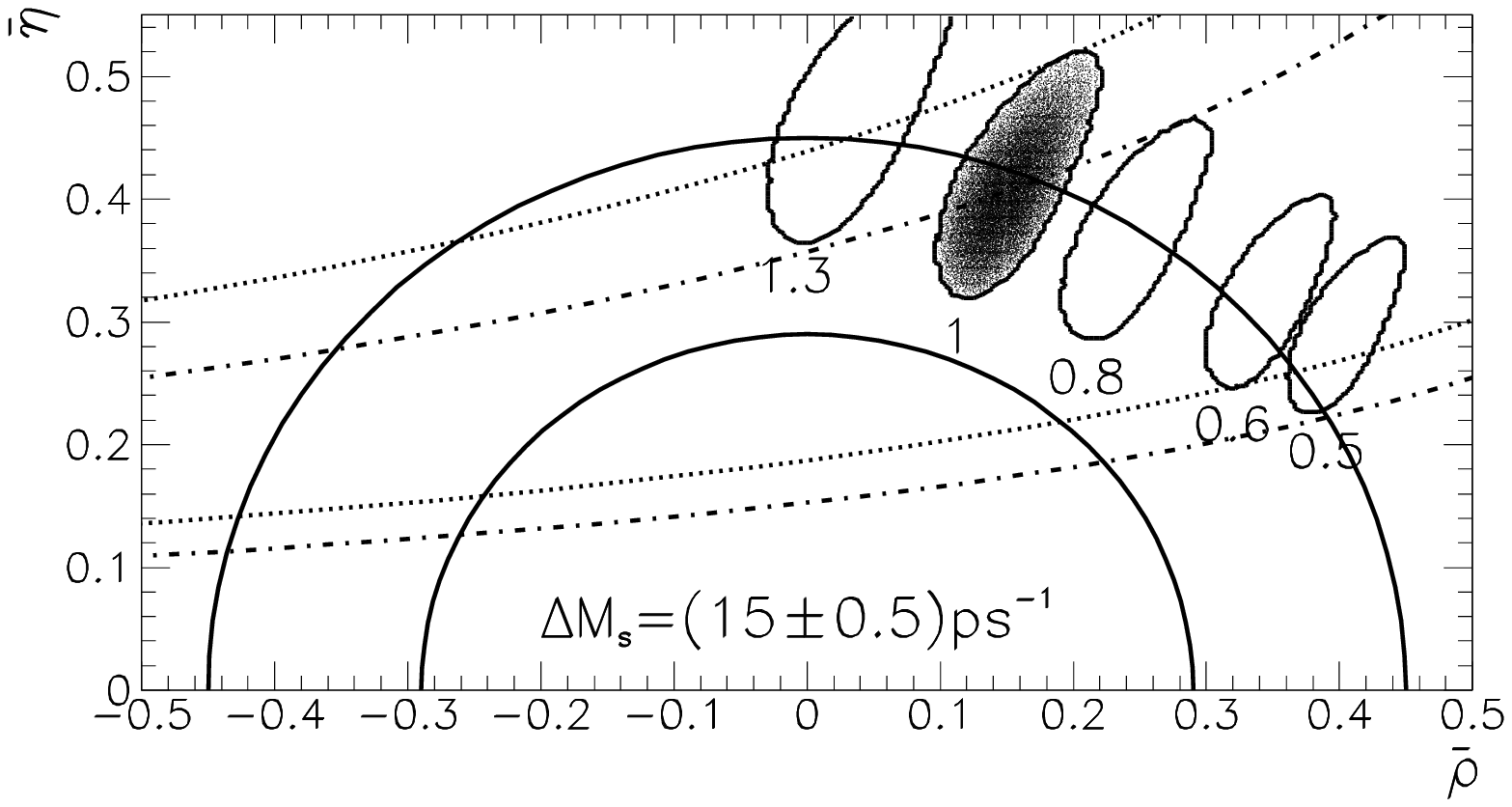,width=\linewidth}}\\
\mbox{\epsfig{file=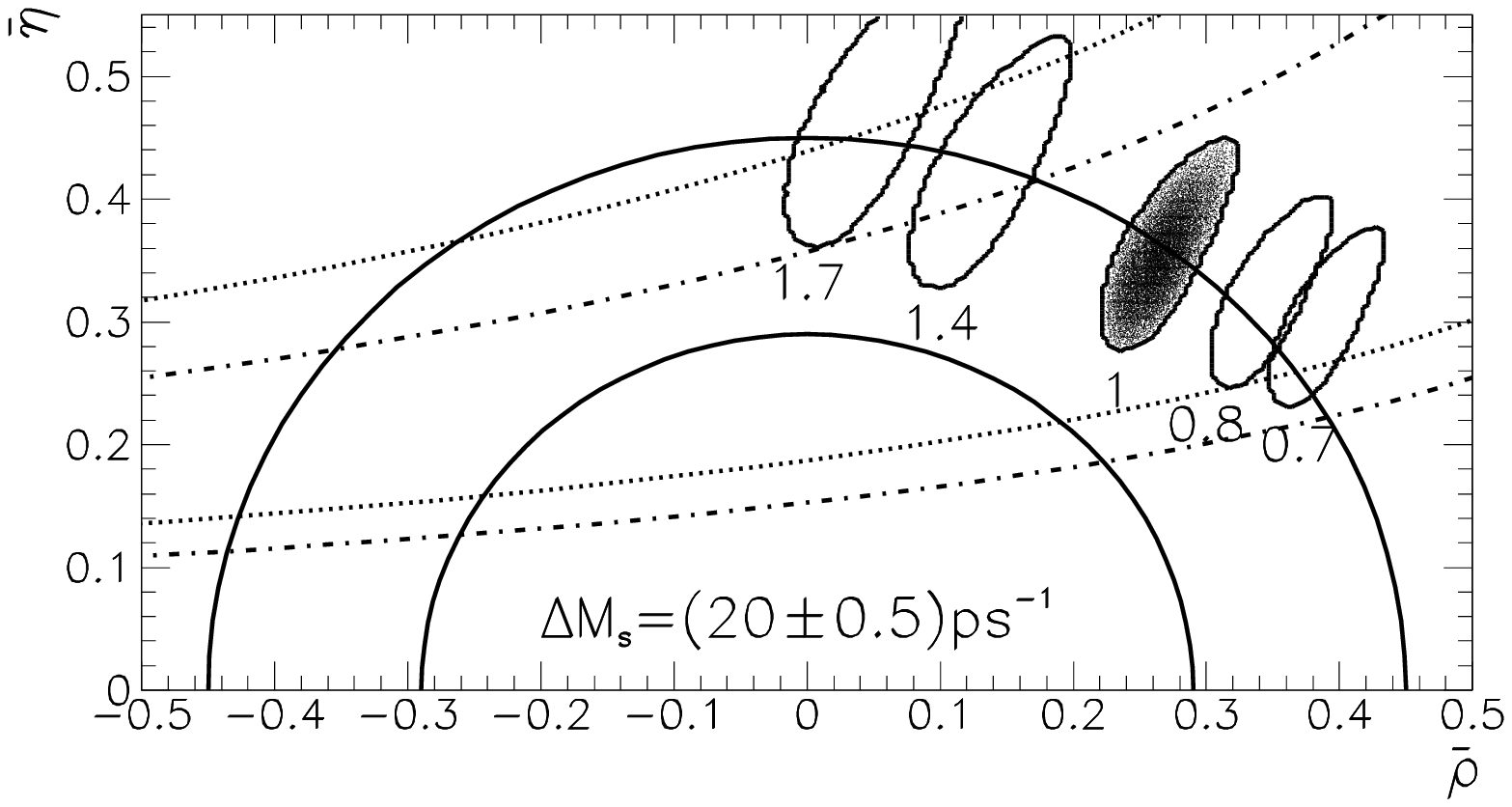,width=\linewidth}}\\
\end{tabular}
\caption{\protect Ranges of $(\bar\rho,\bar\eta)$ allowed at $1\sigma$ for 
  $\Delta M_s =(15.0\pm 0.5)/$ps (upper panel) and $(20.0\pm 0.5)/$ps
  (lower panel) for $\sin2\beta_{\rm ut}=0.78\pm0.08$ and different
  values of $F^s/F^d$ (marked in the figures). Black spots correspond
  to $F^s/F^d=1$.  Dotted (dash dotted) lines show the constraint on
  $(\bar\rho,\bar\eta)$ from $\varepsilon_K$ (eq. (\ref{eqn:epsK}))
  for $F^\varepsilon=F_{\rm SM}$ (for $F^\varepsilon=1.3F_{\rm SM}$ in
  the upper and for $F^\varepsilon=1.5F_{\rm SM}$ in the lower panels,
  respectively) Solid semicircles mark the range of
  $R_b\equiv\sqrt{\bar\rho^2+\bar\eta^2}$ allowed by
  $|V_{ub}/V_{cb}|$.
\label{fig:blobs}}
\end{center}
\end{figure}

Figure \ref{fig:blobs} showing constraints from different experimental
data in the $(\bar\rho,\bar\eta)$ plane allows to discuss the value of
$V_{td}$ in the two scenarios: MSSM with small and large $\tan\beta$
as a function of measured in the future value of $\Delta M_s$.
($R_t\propto|V_{td}|$ equals the length of the line connecting a given
point in the $(\bar\rho, \bar\eta)$ plane with the point $(1,0)$.)

In the MFV-type MSSM with small $\tan\beta$, and also in the SM,
$F^s/F^d=1$ and $\bar\rho$ and $\bar\eta$ are bound to lie inside the
black spots in figure \ref{fig:blobs} which are compatible (for
$\sin2\beta_{\rm ut}\simlt0.78$) with the constraints imposed on $R_b$
by the value of $|V_{ub}/V_{cb}|$. Therefore, $V_{td}$ determined from
$|V_{ub}/V_{cb}|$ ($\propto R_b$), $\Delta M_s/\Delta M_d$ ($\propto
R_t$) and the asymmetry measured in the $B\rightarrow\psi K_S$ decay
($=\sin2\beta_{\rm ut}$) in the MFV-type MSSM and in the SM is the
same: $|V_{td}|=(7.75-9.5)\times10^{-3}$ ($R_t=0.90-0.99$) for $\Delta
M_s=(15\pm0.5)/$ps and $|V_{td}|=(6.7-8.2)\times10^{-3}$
($R_t=0.78-0.85$) for $\Delta M_s=(20\pm0.5)/$ps. Taking into account
the constraint imposed on $\bar\rho$ and $\bar\eta$ by $\varepsilon_K$
does not change anything for $\Delta M_s$ close to 20/ps (the shaded
region lies entirely between the two $\varepsilon_K$ hyperbolae even
for $F^\varepsilon=1.5$). On the other hand, if the value of $\Delta
M_s$ is close to its present lower limit of 15/ps, it follows from
(\ref{eqn:fslim}) that $F^\varepsilon=F^d=F^s$ must be smaller than
$\approx1.3$ (MSSM parameters leading to $F^\varepsilon$ bigger than
1.3 are excluded).  \footnote{Note that this puts severe constraints
  on the scenario with $\tan\beta<1$: stops, charginos and $H^+$ would
  have to be very heavy in order their contribution to $B^0_s$-$\bar
  B^0_s$ mixing described by $F^s$ be sufficiently suppressed.}  Only
for $F^\varepsilon$ close to 1.3 can the upper edge of allowed $R_t$
values (and therefore of $|V_{td}|$) determined from the fit to the
data be slightly lower than in the SM. We conclude therefore that from
the practical point of view the value of $V_{td}$ in the SM and in the
MFV type MSSM is the same. Note also, that in this scenario the
factors $F_{B_s}\sqrt{\hat B_{B_s}}$ and $F_{B_d}\sqrt{\hat B_{B_d}}$
are positively correlated in the sense that, for fixed $\Delta M_s$,
bigger values of $F^\varepsilon=F^d=F^s$ require both these factors to
assume simultaneously values from the lower parts of their respective
ranges obtained from lattice calculations.

In the MSSM with $\tan\beta\sim50$ $F^\varepsilon=F^d=F_{\rm SM}$ and
$F^s/F^d=F^s/F_{\rm SM}<1$. The absolute bound (\ref{eqn:fslim}) does
not allow for $|F^s/F^d|<0.5$ for $\Delta M_s=15$/ps but the
inspection of the upper panel of figure \ref{fig:blobs} shows that the
combination of constraints imposed on $\bar\rho$ and $\bar\eta$ by
$\varepsilon_K$ (dotted lines) and $R_b$ (solid semicircles) excludes
also those MSSM parameters for which $|F^s/F^d|=|F^s/F_{\rm
  SM}|\simlt0.55$. Similarly, for $\Delta M_s=20$/ps the bound
(\ref{eqn:fslim}) gives $|F^s/F^d|>0.69$ whereas $\varepsilon_K$ and
$R_b$ require $|F^s/F^d|\simgt0.75$. For values of $|F^s/F^d|$ at the
lower edge of the allowed range the value of $|V_{td}|$ extracted from
$\Delta M_s/\Delta M_d$ is smaller than in the SM (for example, for
$\Delta M_s=(15\pm0.5)/$ps and $|F^s/F^d|=0.6$ or for $\Delta
M_s=(20\pm0.5)/$ps and $|F^s/F^d|=0.8$
$|V_{td}|=(6.0-7.3)\times10^{-3}$). Note however, that for
$-1>F^s/F^d>-1.3$ for $\Delta M_s=(15\pm0.5)/$ps ($-1>F^s/F^d>-1.76$
for $\Delta M_s=(20\pm0.5)/$ps) the value of $|V_{td}|$ can be bigger
than in the SM. Of course, large departures of $|F^s/F^d|$ from $1$
discussed here are compatible with $\Delta M_s$ and $\Delta M_d$
separately provided the lattice factors $F_{B_s}\sqrt{\hat B_{B_s}}$
and $F_{B_d}\sqrt{\hat B_{B_d}}$ assume appropriate values (which
however remain within their respective uncertainties). The correlation
of $F_{B_s}\sqrt{\hat B_{B_s}}$ and $F_{B_d}\sqrt{\hat B_{B_d}}$ is
again positive (although weaker than in the previous case): smaller 
$F^s/F^d=F^s/F_{\rm SM}$ requires bigger
$F_{B_s}\sqrt{\hat B_{B_s}}$ (to reproduce $\Delta M_s$) and leads to
smaller value of $|V_{td}|$ which in turn calls for bigger
$F_{B_d}\sqrt{\hat B_{B_d}}$ to reproduce $\Delta M_d$.

\section{Impact of the scalar penguins}

As has been pointed out in ref. \cite{ISRE}, the reliable calculation
of the flavour changing neutral Higgs boson couplings in the MSSM
requires resummation of the $\tan\beta$ enhanced terms. Furthermore,
as demonstrated in refs. \cite{DEGAGI,CAGANIWA} there are also
$\tan\beta$ enhanced corrections to the couplings of the charged Higgs
and Goldstone bosons which affect the box diagram contribution of
these particles to the Wilson coefficients of the effective
Hamiltonian (\ref{eqn:heff}). Technical details and systematic study of 
all these refinements can be found in \cite{BUCHROSL2} and will be not 
discussed here. They are however included in the numerical results 
presented below.

\begin{figure}[htbp]
\begin{center}
\epsfig{file=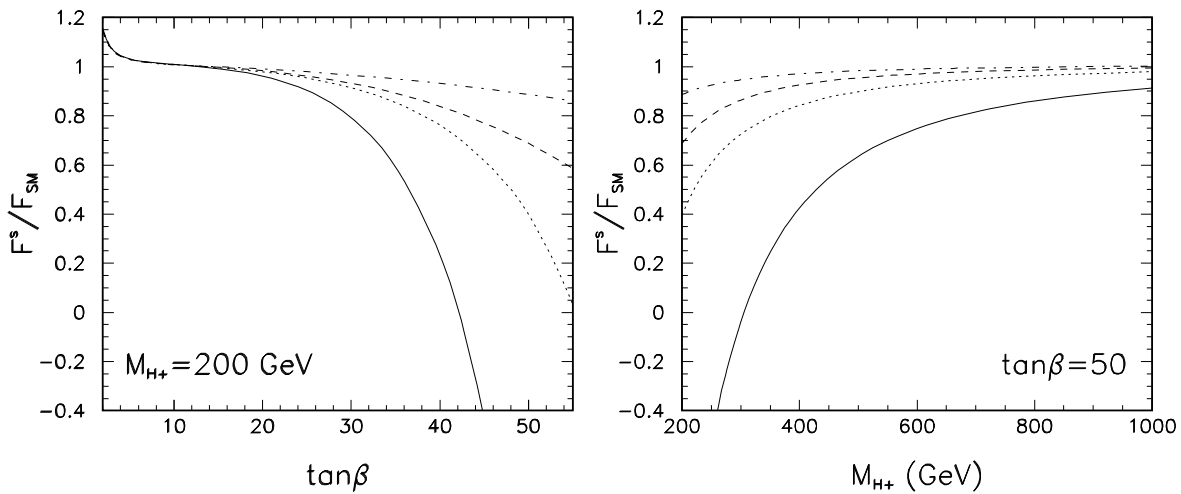,width=\linewidth}
\end{center}
\caption{$F^s/F_{\rm SM}$ as a function of $\tan\beta$ and $M_{H^+}$
  for the lighter chargino mass 750 GeV and $|M_2/\mu|=1$. Solid and
  dashed dashed lines correspond to stop masses (in GeV) (500,850)
  whereas dotted and dot-dashed lines to (600, 750). The mixing angle
  between the two stops is $|\theta_{\tilde t}|=10^o$. Solid and
  dotted (dashed and dot-dashed) lines correspond to $\mu<0$ ($\mu>0$)
  and the stop mixing angle has the sign opposite to that of $\mu$.
  $m_{\tilde g}=3M_2$ and the right sbottom mass is 800 GeV.}
\label{fig:1fs}
\end{figure}

The role of the scalar penguin induced flavour changing neutral Higgs
boson couplings is twofold. Firstly, for $\tan\beta\simgt30$ a big
portion of the MSSM parameter space (the bigger the higher is the
lower experimental limit on $\Delta M_s$) in which the parameter $A_t$
is large (and hence the mixing of left and right top squarks is
substantial) is excluded by the bound (\ref{eqn:fslim}) and its
refinement related to constraints on $\bar\rho$ and $\bar\eta$ from
$\varepsilon_K$ and $|V_{ub}/V_{cb}|$ discussed in the preceding
section.  Typical dependence of $F^s/F_{\rm SM}$ on the MSSM
parameters is shown in figure \ref{fig:1fs}. For $\mu>0$ the
resummation of $\tan\beta$ enhanced terms mentioned above increases
\cite{ISRE} the value of $F^s/F_{\rm SM}$ (i.e. suppresses the
negative contribution of the flavour changing couplings of neutral
Higgs bosons) compared to the naive one-loop calculation of ref.
\cite{BUCHROSL,BOEWKRUR}. For $\mu<0$, however, the effects of the
flavour changing couplings are enhanced by the resummation. The
parameters in figure \ref{fig:1fs} has been chosen so that $\mu A_t$
has always the sign \cite{CAGANIWA} which allows for cancellation of
the $tH^+$ and chargino-stop contributions to the amplitude of the
$\bar B\rightarrow X_s\gamma$ decay.

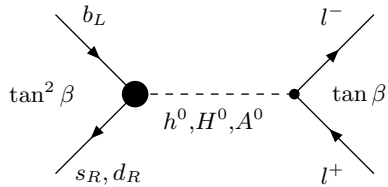
\begin{figure}[htbp]
\begin{center}
\begin{picture}(160,80)(0,0)
\ArrowLine(20,70)(50,40)
\ArrowLine(50,40)(20,10)
\Vertex(50,40){5}
\ArrowLine(140,10)(110,40)
\ArrowLine(110,40)(140,70)
\DashLine(50,40)(110,40){3}
\Vertex(110,40){2}
\Text(80,30)[]{\small $h^0$,$H^0$,$A^0$}
\Text(35,70)[]{\small $b_L$}
\Text(40,10)[]{\small $s_R,d_R$}
\Text(125,70)[]{\small $l^-$}
\Text(125,10)[]{\small $l^+$}
\Text(15,40)[]{\small $\tan^2\beta$}
\Text(135,40)[]{\small $\tan\beta$}
\end{picture}
\end{center}
\caption{Flavour changing neutral Higgs boson couplings contribution
  to the amplitude of the $B^0_{s,d}\rightarrow l^+l^-$ decays.}
\label{fig:bmumu}
\end{figure}

Secondly, the same flavour changing neutral Higgs boson couplings
which (through the double penguin diagrams) affect the $B^0_s$-$\bar
B^0_s$ mixing has been found \cite{BAKO,CHSL,BOEWKRUR} to totally
dominate for $\tan\beta\simgt30$ amplitudes of the decays $B^0_{s,d}
\rightarrow \mu^+\mu^-$. Calculating the diagram shown in fig.
\ref{fig:bmumu} one finds \cite{CHSL,BOEWKRUR}
\begin{eqnarray}
{\cal A}(B^0_q\rightarrow\mu^+\mu^-)=
\bar u(k_1)\left(b+a\gamma^5\right)v(k_2)\label{eqn:bmumuamp}
\end{eqnarray}
where $q=s$ or $d$, $u(k_1)$, $v(k_2)$ are spinors of the final state
leptons, and (without resummations, with $M_H\approx M_A$ etc.)
\begin{eqnarray}
a=b=-V_{tb}^\ast V_{tq}m_lF_{B_q}{G_F\alpha_{\rm em}\over8\sqrt2s^2_W}
{M^2_{B_q}\over M^2_A}{m^2_t\over M^2_W} X_{tC}\tan^3\beta ~.\nonumber
\end{eqnarray}
Therefore, in the MSSM with large $\tan\beta$ the decay rate behaves
as $BR(B^0_{s,d}\rightarrow\mu^+\mu^-)\propto(\tan^6\beta/M_A^4)$ and
- without additional constraint imposed - could, for
$\tan\beta\simgt50$ and the Higgs bosons not too heavy, even exceed
the present experimental bounds \cite{ALEXAN}
\begin{eqnarray}
BR(B^0_s\rightarrow\mu^+\mu^-)<2.0\times10^{-6}\phantom{aaaa}&&{\rm CDF}
\nonumber\\
BR(B^0_d\rightarrow\mu^+\mu^-)<2.1\times10^{-7}\phantom{aaaa}&&{\rm BaBar}
\label{eqn:Bll_lim}
\end{eqnarray}
That is, the rates predicted in the MSSM could exceed by 3-4 orders of
magnitude those of the SM \cite{BUBU,BUREV,ALMA}: 
\begin{eqnarray}
&&BR(B^0_s\rightarrow\mu^+\mu^-)_{\rm SM}\approx3.5\times10^{-9}
\left({F_{B_s}\over 230 ~{\rm MeV}}\right)^2\nonumber\\
&&BR(B^0_d\rightarrow\mu^+\mu^-)_{\rm SM}\approx1.4\times10^{-10}
\left({F_{B_d}\over 200 ~{\rm MeV}}\right)^2
\left({|V_{td}|\over0.009}\right)^2\nonumber
\end{eqnarray}

However, as we have discussed above, for light $A^0$ the magnitude of
the flavour changing scalar couplings $b_R A^0 s_L$ and $b_R H^0 s_L$
(and, hence, also of the couplings $b_R A^0 d_L$ and $b_R H^0 d_L$,
because in the GMFV MSSM they are proportional to the former ones) is
strongly constrained by the condition (\ref{eqn:fslim}). Therefore one
can expect that also the contribution of the neutral Higgs boson
exchange shown in figure \ref{fig:bmumu} to the amplitudes of the
$B^0_{s,d}\rightarrow\mu^+\mu^-$ decays is bounded by the condition
(\ref{eqn:fslim}). In other words, 
the lower limit on
$\Delta M_s$ should put the upper bound on the possible values of
$BR(B^0_{s,d}\rightarrow\mu^+\mu^-)$ predicted in the MSSM.

\begin{figure}[htbp]
\begin{center}
\epsfig{file=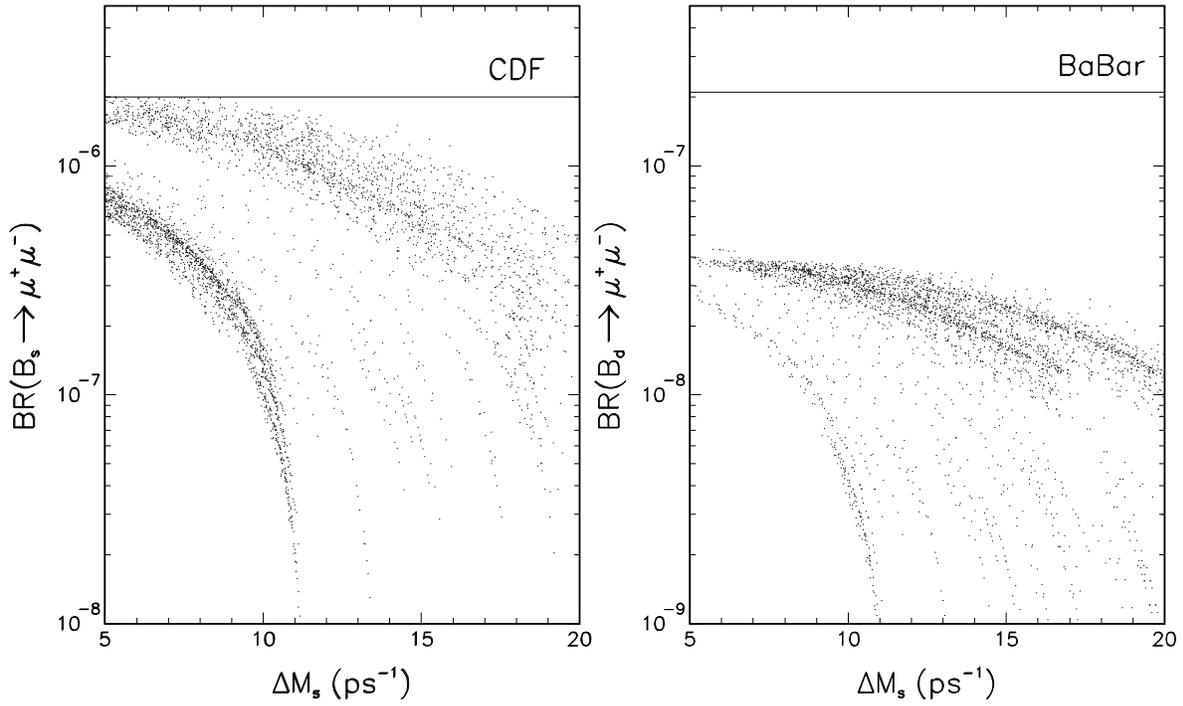,width=\linewidth}
\end{center}
\caption{\protect Correlation of $BR(B^0_s\rightarrow\mu^+\mu^-)$ (left 
  panel) and $BR(B^0_d\rightarrow\mu^+\mu^-)$ (right panel) with $\Delta M_s$
  in the GMFV-type MSSM for $\tan\beta=50$ and $M_A=200$ GeV. In this case
  all points for which $F^s/F_{\rm SM}<-0.52$ (so that $\Delta M_s>15/$ps) 
  give $BR(B^0_s\rightarrow\mu^+\mu^-)$ above the CDF bound 
  (\ref{eqn:Bll_lim}) and have been discarded.}
\label{fig:corel}
\end{figure}

Figure \ref{fig:corel}, shows the correlation of the predicted values
of $BR(B^0_{s,d}\rightarrow\mu^+\mu^-)$ and $\Delta M_s$ for a sample
of the MSSM parameters for $M_A=200$ GeV and $\tan\beta=50$. In the
case of the $BR(B^0_d\rightarrow\mu^+\mu^-)$ we have determined the
value of $|V_{td}|$ consistently, that is we have scanned over the
Wolfenstein parameters $\lambda$, $A$, $\bar\rho$ and $\bar\eta$ as
well as over the nonperturbative parameters $F_{B_q}\sqrt{\hat
  B_{B_q}}$ and computed the rate only for those $\lambda$, $A$,
$\bar\rho$ and $\bar\eta$ for which $\epsilon_K$, $\Delta M_d$,
$\sin2\beta_{\rm ut}$, $|V_{ub}/V_{cb}|$ assumed acceptable values. We
have also excluded all points for which the rate of the $\bar
B\rightarrow X_s\gamma$ is unacceptable.

The upper bounds on $BR(B^0_{s,d}\rightarrow\mu^+\mu^-)$ are clearly seen 
in figure \ref{fig:corel}. For $\tan\beta=50$ and $M_A=200$ GeV all 
points for which $F^s/F_{\rm SM}<-0.52$ (so that $\Delta M_s>15/$ps) give 
$BR(B^0_s\rightarrow\mu^+\mu^-)$ above the CDF bound (\ref{eqn:Bll_lim}) 
and excluding also points for which $\Delta M_s<15/$ps we see, that
$BR(B^0_s\rightarrow\mu^+\mu^-)<10^{-6}$ and
$BR(B^0_d\rightarrow\mu^+\mu^-)<3\times10^{-8}$. Points for which 
$F^s/F_{\rm SM}<-0.52$ can survive for smaller values of 
$\tan\beta$ and/or heavier CP-odd scalar $A^0$ (note that 
$BR\propto\tan^6\beta/M^4_A$ whereas $\Delta M_s\propto|F^s|
\propto\tan^4\beta/M^2_A$). In this case however both, $\Delta M_s$ and
$BR(B^0_{s,d}\rightarrow\mu^+\mu^-)$ are entirely dominated by the  
contributions of the scalar penguins and it is easy to estimate that
whenever $BR(B^0_s\rightarrow\mu^+\mu^-)$ is below the CDF bound 
(\ref{eqn:Bll_lim}), $BR(B^0_d\rightarrow\mu^+\mu^-)\simlt6\times10^{-8}$,
i.e. it is below the  BaBar bound (\ref{eqn:Bll_lim}).

We conclude that the MSSM parameter space in which the parameter $A_t$
is not unnaturally big (that is, $A_t\simlt M_{\rm SUSY}$) is more strongly 
constrained by the lower limit on $\Delta M_s$ than by the non-observation 
of the $B^0_{s,d}\rightarrow\mu^+\mu^-$ decays in CDF and BaBar. In 
particular the bound $BR(B^0_d\rightarrow\mu^+\mu^-)<3\times10^{-8}$ holds.
For parameters such that $F^s/F_{\rm SM}<-0.52$ (larger  $A_t$) there is a 
weaker upper bound $BR(B^0_d\rightarrow\mu^+\mu^-)\simgt6\times10^{-8}$

\section{Flavour violation in squark mass matrices}

\begin{figure}[htbp]
\begin{center}
\begin{picture}(350,80)(0,0)
\ArrowLine(130,20)(150,20)
\ArrowLine(150,20)(200,20)
\ArrowLine(200,20)(220,20)
\ArrowLine(220,70)(200,70)
\ArrowLine(200,70)(150,70)
\ArrowLine(150,70)(130,70)
\DashLine(150,70)(150,20){3}
\DashLine(200,20)(200,70){3}
\Line(148,43)(152,47)
\Line(148,47)(152,43)
\Line(198,43)(202,47)
\Line(198,47)(202,43)
\Text(120,15)[]{$d_I$}
\Text(230,15)[]{$d_J$}
\Text(120,65)[]{$d_J$}
\Text(230,65)[]{$d_I$}
\Text(175,9)[]{$\tilde g$}
\Text(175,59)[]{$\tilde g$}
\Text(163,38)[]{$\tilde d$}
\Text(190,38)[]{$\tilde d$}
\Text(135,42)[]{$\delta_{XY}^D$}
\Text(215,42)[]{$\delta_{XY}^D$}
\Text(255,45)[]{$\propto\alpha_s^2$}
\end{picture}
\end{center}
\caption{Contribution of gluino-squark box diagrams neutral meson mixing. 
  Crosses denote mass insertions.}
\label{fig:glbox}
\end{figure}
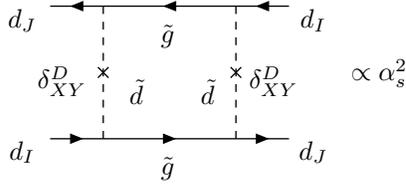

In supersymmetric extension of the SM, flavour and CP violation can
originate also in the sfermion sector. In general, the $6\times6$ mass
squared matrices of left- and right-chiral sfermions of the same
electric charge have the form\footnote{Except for sneutrinos whose
  mass squared matrix consists of the LL $3\times3$ block only.}
\begin{eqnarray}
{\cal M}^2_Q=
\left(\matrix{\left(M^Q_{LL}\right)^2 & \left(M^Q_{LR}\right)^2
\cr \left(M^Q_{RL}\right)^2 & \left(M^Q_{RR}\right)^2}\right)
\phantom{aaa} Q=U,D,L\nonumber
\end{eqnarray}
where $\left(M^Q_{LL}\right)^2$ etc. are $3\times3$ submatrices. If
the latters are not diagonal in the so-called superCKM basis, in which
quark mass matrices are diagonal, then their off-diagonal entries
generate flavour changing neutral currents.  For example, large,
$\propto\alpha_s^2$, contributions to $K^0$-$\bar K^0$ or
$B^0_{s,d}$-$\bar B^0_{s,d}$ mixing are then induced by the gluino box
diagrams shown in figure \ref{fig:glbox}. In this figure the
off-diagonal entries of matrices $\left(M^Q_{XY}\right)^2$ are treated
as additional interactions (the so-called mass insertion approximation
\cite{GAGAMASI,CIETAL,MIPORO}). As these contributions are not
proportional to the CKM matrix factors the effective Hamiltonian
(\ref{eqn:heff}) for $|\Delta F|=2$ transitions has to be now written
as
\begin{eqnarray}
{\cal H}_{\rm eff}=\sum_X  C_X{\cal O}_X\nonumber
\end{eqnarray}
where $C_X$ are the Wilson coefficients computed in the MSSM and
${\cal O}_X$ are the same four-quark operators as in eq.
(\ref{eqn:heff}). Assuming for definiteness that sparticle masses are
of the order of $M_{\rm SUSY}=500$ GeV, and taking into account the
QCD RG running of $C_X$ between $M_{\rm SUSY}$ and the hadronic scale
as well as matrix elements of the operators ${\cal O}_X$ between the
meson states in the manner described in \cite{BUJAUR} one obtains for
the supersymmetric contribution to the $K^0$-$\bar K^0$ transition
amplitude:
\begin{eqnarray}
&&\langle\bar K^0|{\cal H}_{\rm eff}|K^0\rangle\approx M_KF^2_K
\left[0.15\left(C^{\rm VLL}_{\rm SUSY}+C^{\rm VRR}_{\rm SUSY}\right)
-6.0\left(C^{\rm SLL}_{\rm SUSY}+C^{\rm SRR}_{\rm SUSY}\right)\right.
\nonumber\\
&&\phantom{aaaaaaaa}\left.
-11.5\left(C^{\rm TL}_{\rm SUSY}+C^{\rm TR}_{\rm SUSY}\right)
-13.84 ~C^{\rm VLR}_{\rm SUSY}+22.48 ~C^{\rm SLR}_{\rm SUSY}\right]
\label{eqn:KKtrans}
\end{eqnarray}
where we have used $\alpha_s(M_Z)=0.1185$.  The large numerical
factors\footnote{Some uncertainties of order few percent in these
  numbers are due to the uncertainties of the $B_K^X$ factors
  parameterizing matrix elements of the operators ${\cal O}_X$ for
  X=SLL, SRR, VLR, SLR, TL, TR. \cite{AL}} in the second line
originate from the RG running and from the chiral enhancement factor
$(M_K/(m_s+m_d))^2\approx18$ for $m_s(2 ~{\rm GeV})=110$ MeV. For the
supersymmetric contribution to the $B^0_q$-$\bar B^0_q$ transition
amplitude one has to replace in eq.  (\ref{eqn:KKtrans}) $M_KF^2_K$ by
$M_{B_q}F^2_{B_q}$ ($q=d$ or $s$) and the numbers in the square
bracket by: $0.24$, $-0.49$, $-0.94$, $-0.97$ and $1.27$,
respectively. Note, that there is no chiral enhancement in this case
as $(M_{B_q}/(m_b+m_d))^2\approx1.65$.

Using the standard formulae
\begin{eqnarray}
&&\Delta M_K = 2~{\rm Re} ~\langle\bar K^0|{\cal H}_{\rm eff}|K^0\rangle
\nonumber\\
&&\varepsilon_K = {e^{i\pi/4}\over\sqrt2\Delta M_K}
                  {\rm Im} ~\langle\bar K^0|{\cal H}_{\rm eff}|K^0\rangle
\nonumber\\
&&\Delta M_q = 2~|\langle\bar B^0_q|{\cal H}_{\rm eff}|B_q^0\rangle|
\nonumber
\end{eqnarray}
and plugging in numbers one finds
\begin{eqnarray}
&&\Delta M_K = 
3.87\times10^4~{\rm Re} ~\left(M^2_{\tilde q}\times[\dots]\right)
\left({1 ~{\rm TeV}\over M^2_{\tilde q}}\right)^2 ~{\rm ps}^{-1}\nonumber\\
&&\varepsilon_K = -2.58\times10^6 ~{\rm Im} ~\left(M^2_{\tilde q}\times
[\dots]\right)
\left({1 ~{\rm TeV}\over M^2_{\tilde q}}\right)^2e^{i\pi/4}\label{eqn:num}\\
&&\Delta M_d = 6.45\times10^5~{\rm Re} ~\left(M^2_{\tilde q}\times
[\dots]\right)
\left({1 ~{\rm TeV}\over M^2_{\tilde q}}\right)^2 ~{\rm ps}^{-1}\nonumber\\
&&\Delta M_s = 8.78\times10^5~{\rm Re} ~\left(M^2_{\tilde q}\times
[\dots]\right)
\left({1 ~{\rm TeV}\over M^2_{\tilde q}}\right)^2 ~{\rm ps}^{-1}\nonumber
\end{eqnarray}
where $M^2_{\tilde q}$ is some average mass of squarks and $[\dots]$ denote 
the content of square brackets from eq. (\ref{eqn:KKtrans}) appropriate for 
a given transition.

In the lowest order in the mass insertion approximation each of the
Wilson coefficients $C_X$ for $|\Delta F|=2$ transitions like
$K^0$-$\bar K^0$ and $B^0_q$-$\bar B^0_q$ can be represented as a
product of a function of $M^2_{\tilde q}$ and $m_{\tilde g}$ and of
two mass insertions defined as \cite{GAGAMASI,CIETAL,MIPORO}
\begin{eqnarray}
(\delta^D_{XY})^{JI}={[(M^D_{XY})^2]^{JI}\over M^2_{\tilde q}}\nonumber
\end{eqnarray}
where $X,Y=L,R$, and the indices $J,I$ label generations. Neglecting 
contributions other than those generated by gluino exchanges, one has
\begin{eqnarray}
C^{\rm VLL}_{\rm SUSY}=a ~\alpha_s^2\left[(\delta^D_{LL})^{JI}\right]^2,
\phantom{aaa}
C^{\rm VRR}_{\rm SUSY}=b ~\alpha_s^2\left[(\delta^D_{RR})^{JI}\right]^2,
\nonumber\\
C^{\rm SLR}_{\rm SUSY}\propto \alpha_s^2\left[
a^\prime ~(\delta^D_{LL})^{JI}(\delta^D_{RR})^{JI}+
b^\prime ~(\delta^D_{LR})^{JI}(\delta^D_{RL})^{JI}\right]
\phantom{a}
\label{eqn:cstructure}
\end{eqnarray}
etc. \cite{CIETAL}, where $JI=21$, $31$ and $32$ for  $K^0$-$\bar K^0$,
$B^0_d$-$\bar B^0_d$ and $B^0_s$-$\bar B^0_s$ transitions, respectively.

Comparison of the numbers in eqs. (\ref{eqn:num}) with the
experimental values: $\Delta M_K=0.0053/$ps,
$\varepsilon_K=2.28\times10^{-3}$ and $\Delta M_d=0.496/$ps
illustrates the so-called supersymmetric flavour and CP problem:
taking into account that the dimensionless factors $M^2_{\tilde
  q}\times[\dots]$ in eqs. (\ref{eqn:num}) are $\sim{\cal O}(1)$, it
is clear that the typical contribution to $\Delta M_K$,
$\varepsilon_K$, $\Delta M_{B_q}$ and to many other measured
quantities like $\varepsilon^\prime/\varepsilon$, $BR(\bar
B\rightarrow X_s\gamma)$ \cite{OTHERS,MIPORO,IS}, etc. in the MSSM
with the flavour and CP violation in squark mass matrices is several
orders of magnitude too big. Any theory of supersymmetry breaking has
to face the problem of explaining the smallness of the mass insertions
$\delta_{XY}^Q$.

Adopting the rough criterion that the gluino contribution alone to
$|\Delta M_K|$ and $|\varepsilon_K|$ should not exceed the
experimental values of these quantities (and barring possible
cancellation between different mass insertions) one obtains for small
and moderate values of $\tan\beta$ the limits shown in the middle
column of table \ref{tab:limits}.  For comparison in the first column
we show the limits obtained in the paper \cite{CIETAL}. The
differences stem from slightly different treatment of the NLO QCD RG
evolution and of the matrix elements of the operators involved (our
approach is based on ref. \cite{BUJAUR}) but are inessential for the
order of magnitude estimates of the limits.

\begin{table}[thb]
\caption[]{Upper limits on mass insertions obtained from $\varepsilon_K$ for 
$M_{\tilde q}=500$ GeV. The limits scale approximately as  $M_{\tilde q}^2$.
$x\equiv(m_{\tilde g}/M_{\tilde q})^2$. Limits on 
$\left[ (\delta^D_{12})_{RR}^2\right]$ are the same as for
$\left[ (\delta^D_{12})_{LL}^2\right]$.
As follows from numbers in the first two of eqs. (\ref{eqn:num}), 
the corresponding limits on real parts of the product of insertions
are simply $12.5$ times weaker than those given below. (This simple
rule is not satisfied by the numbers quoted in ref. \cite{CIETAL}.)\\
\label{tab:limits}}
\begin{center}
\begin{tabular}{||c|c|c|c||}  \hline \hline
& ref. \cite{CIETAL} & $\tilde g$ & $\tilde g$ \\
&          & low $\tan\beta$     & $\tan\beta=50$ \\ 
\hline
$x$ & \multicolumn{3}{c||}
{$\sqrt{|{\rm Im}\left[ (\delta^D_{12})_{LL}^2\right]|} $}
\\
\hline
0.3
& $2.9\times 10^{-3}$
& $2.7\times 10^{-3}$
& $2.5\times 10^{-3}$
\\
1.0
& $6.1\times 10^{-3}$
& $6.0\times 10^{-3}$
& $1.7\times 10^{-3}$
\\
4.0
& $1.4\times 10^{-2}$
& $1.5\times 10^{-2}$
& $2.0\times 10^{-3}$
\\
9.0
& $--$
& $1.4\times 10^{-2}$
& $2.4\times 10^{-3}$
\\
\hline 
$x$
& \multicolumn{3}{c||}
{$\sqrt{|{\rm Im}\left[(\delta^D_{12} )_{LR}^2\right]|}
\qquad (|(\delta^D_{12})_{LR}|\gg |(\delta^D_{12})_{RL}|)$}
\\
\hline
0.3
& $3.4\times 10^{-4}$
& $2.5\times 10^{-4}$
& $2.2\times 10^{-4}$
\\
1.0
& $3.7\times 10^{-4}$
& $2.9\times 10^{-4}$
& $2.2\times 10^{-4}$
\\
4.0
& $5.2\times 10^{-4}$
& $4.2\times 10^{-4}$
& $2.5\times 10^{-4}$
\\
4.0
& $--$
& $6.5\times 10^{-4}$
& $6.9\times 10^{-4}$
\\
\hline
$x$
& \multicolumn{3}{c||}{$\sqrt{|{\rm Im}\left[ (\delta^D_{12} )_{LL}
(\delta^D_{12})_{RR}\right]|} $}
\\
\hline
0.3
& $1.1\times 10^{-4}$
& $8.2\times 10^{-5}$
& $8.0\times 10^{-5}$
\\
1.0
& $1.3\times 10^{-4}$
& $9.5\times 10^{-5}$
& $9.2\times 10^{-5}$
\\
4.0
& $1.8\times 10^{-4}$
& $1.4\times 10^{-4}$
& $1.3\times 10^{-4}$
\\
9.0
& $--$
& $1.9\times 10^{-4}$ 
& $1.8\times 10^{-4}$ 
\\
\hline\hline 
\end{tabular} 
\end{center}
\end{table}

It is interesting to note, that because $\varepsilon_K$ puts stringent
bounds only on imaginary parts of products of two mass insertions,
bounds on almost real and almost imaginary mass insertions are
provided only by $\Delta M_K$ and are order of magnitude weaker,
although such a conspiracy seems not very natural. Stronger absolute
bound on the imaginary part of the mass insertion itself exist only
for $(\delta^D_{LR})^{12}$ and follows from
$\varepsilon^\prime/\varepsilon$: $|{\rm
  Im}(\delta^D_{LR})^{12}|\simlt10^{-5}$ \cite{GAGAMASI,MAMU}.

For mass insertions generating transitions between the third and the
first two generations of quarks only much weaker bounds are available.
Limits on $(\delta^D_{XY})^{13}$ insertions from the gluino box
contribution to $\Delta M_d$ have been derived recently in ref.
\cite{MASI}. Similar limits on $(\delta^D_{XY})^{23}$ insertions will
become available once $\Delta M_s$ is measured. At present however,
stringent bounds from the $\bar B\rightarrow X_s\gamma$ decay exist
only for the insertions $(\delta^D_{LR})^{23}$ and
$(\delta^D_{RL})^{23}$: $|(\delta^D_{LR})^{23}|<0.07\times(M_{\tilde
  q}/1 ~{\rm TeV})$. The remaining insertions are bounded rather
weakly \cite{MIPORO,IS}.

For large $\tan\beta$ the standard analysis of bounds on mass
insertions derived from $|\Delta F|=2$ transitions based on gluino box
diagrams of figure \ref{fig:glbox} is not sufficient. Scalar flavour
changing neutral Higgs boson couplings can generate additional
contributions $\propto\tan^4\beta$ to the Wilson coefficients
$C^{SLR}$ through the double penguin diagrams of figure \ref{fig:2pg}
(contributions to $C^{SLL}$, and $C^{SRR}$ Wilson coefficients are
suppressed because of the mutual cancellation of $H^0$ and $A^0$
contributions) and these contributions have to be taken into account.
Dominant source of the flavour changing neutral Higgs boson couplings
in the case of flavour violation in squark mass matrices are the
diagrams b) and c) shown in figure \ref{fig:scpg}. Calculating those
diagrams one gets the couplings
\begin{eqnarray}
{\cal L}= S^0 \overline{d^J_R} \left[X_{RL}^S\right]^{JI} d_L^I + 
S^0 \overline{d^J_L} \left[X_{LR}^S\right]^{JI} d_R^I\phantom{aaa}J\neq I
\label{eqn:fcnc_coupl}
\end{eqnarray}
where the matrix coefficients $X_{RL}^S=(X^S_{LR})^\dagger$ are given by
\begin{eqnarray}
&&\left[X_{RL}^S\right]^{JI}=x^S_d\tan^2\beta{e\alpha_s\over3\pi s_W}
{m_{\tilde g}\mu^\ast\over M^2_{\tilde q}}
\left[\left(\delta^D_{LL}\right)^{IJ}{m_{d_J}\over M_W}+{m_{d_I}\over M_W}
\left(\delta^D_{RR}\right)^{IJ}\right] \nonumber\\
&&\phantom{aaaaaaaa}\times D(m^2_{\tilde g},M^2_{\tilde q})\nonumber
\end{eqnarray}
with $x^S_d=\cos\alpha$, $-\sin\alpha$ and $i\sin\beta$ for 
 $S^0=H^0$, $h^0$ or $A^0$, respectively and $D(a,b)$ some dimensionless
function.

It turns out that even the limits on $(\delta^D_{LL})^{12}$ and
$(\delta^D_{RR})^{12}$ mass insertions are affected by the double
penguin contribution which is significantly enhanced by the big
numerical factor multiplying $C^{SLR}_{\rm SUSY}$ in eq.
(\ref{eqn:KKtrans}).  The effect of double penguin contribution is
seen in table \ref{tab:limits} in the limits on imaginary (and real)
parts of $[(\delta^D_{LL})^{12}]^2$ and $[(\delta^D_{RR})^{12}]^2$
which for $m_{\tilde g} > M_{\tilde q}$ become stronger by one order
of magnitude compared to similar limits for lower $\tan\beta$ values.
That the improvement is seen only for $m_{\tilde g} > M_{\tilde q}$
follows from the fact that the couplings (\ref{eqn:fcnc_coupl}) are
proportional to $m_{\tilde g}$. The limits on
$(\delta^D_{LL})^{12}(\delta^D_{RR})^{12}$ are not improved because,
as is clear from eq. (\ref{eqn:cstructure}), the gluino box
contribution to $C^{SLR}_{\rm SUSY}$ contains already a term
proportional to $(\delta^D_{LL})^{12}(\delta^D_{RR})^{12}$.

In the same manner, bounds on the insertions $(\delta^D_{LL})^{13}$
and $(\delta^D_{RR})^{13}$ (and when $\Delta M_s$ is measured also on
$(\delta^D_{LL})^{23}$ and $(\delta^D_{RR})^{23}$) derived from
$B^0$-$\bar B^0$ mixing \cite{MASI} should also be modified for large
values of $\tan\beta$. We have found, however, that the actual bounds
depend also the chargino box and double penguin contributions and can
not be therefore presented in a simple way.

\begin{figure}[htbp]
\begin{center}
\epsfig{file=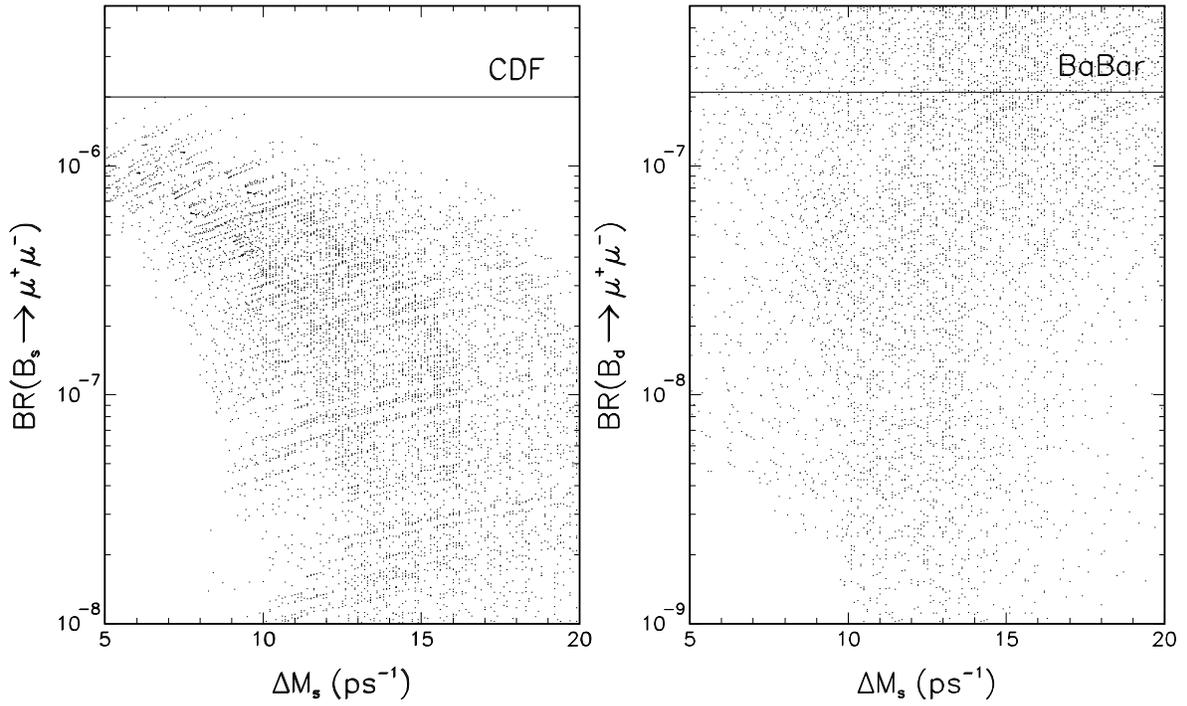,width=\linewidth}
\end{center}
\caption{Correlation of $BR(B^0_s\rightarrow\mu^+\mu^-)$ (left panel)
  and $BR(B^0_d\rightarrow\mu^+\mu^-)$ (right panel) with $\Delta M_s$
  in the MSSM with flavour violation in the squark sector. The single
  nonzero mass insertion $\left(\delta^D_{LL}\right)^{31}$ has been
  varied in the range $(0.01,0.1)$. $\tan\beta=50$ and $M_A=200$ GeV.}
\label{fig:inser}
\end{figure}

Another interesting effect related to the flavour changing couplings
(\ref{eqn:fcnc_coupl}) generated for large $\tan\beta$ by non-zero LL
and/or RR mass insertions is a growing like $\tan^6\beta$ contribution
to the amplitudes of $B^0_s\rightarrow\mu^+\mu^-$ and/or
$B^0_d\rightarrow\mu^+\mu^-$ decays \cite{CHSL}. Calculating the
contribution of the diagram shown in figure \ref{fig:bmumu} with the
couplings (\ref{eqn:fcnc_coupl}) one finds for the coefficients $a$ in
the amplitude (\ref{eqn:bmumuamp})
\begin{eqnarray}
&&a=F_{B_q}m_l {e^2\alpha_s\over12\pi s^2_WM^2_W}
{M^2_B\over M^2_A} \tan^3\beta
\left[{m_{\tilde g}\mu^\ast\over M^2_{\tilde q}}
\left(\delta^D_{LL}\right)^{3q}
+{m_{\tilde g}\mu\over M^2_{\tilde q}}
\left(\delta^D_{RR}\right)^{3q}\right] \nonumber\\
&&\phantom{aaaaa}\times D(m^2_{\tilde g},M^2_{\tilde q})\nonumber
\end{eqnarray}
where $q=d=1$ and $q=s=2$ for $B^0_d$ and $B^0_s$ decays,
respectively, and the coefficient $b$ in the amplitude
(\ref{eqn:bmumuamp}) is given by the similar expression with $+$
changed to $-$ in the square bracket. It has been shown \cite{CHSL},
that for $\tan\beta\sim50$, $M_A\simgt200$ GeV and with mass
insertions of order $0.1$ the branching ratios predicted in the MSSM
can exceed by one or two orders of magnitude the present experimental
limits (\ref{eqn:Bll_lim}). It is however important to check whether
this remains true when all the available constraints are respected,
including the ones imposed by $\Delta M_s$ and $\Delta M_d$ and taking
consistently into account the double penguin contributions to these
quantities. The results of this exercise are shown in figure
\ref{fig:inser} where we show the branching ratios of the decays
$B^0_s\rightarrow\mu^+\mu^-$ and $B^0_d\rightarrow\mu^+\mu^-$ versus
the mass difference $\Delta M_s$ for a sample of the MSSM parameters
varying the insertion $\left(\delta^D_{LL}\right)^{31}$ in the range
$(0.01,0.1)$. All points giving rise to experimentally unacceptable
values of $\Delta M_d$ and/or $BR(\bar B\rightarrow X_s\gamma)$ have
been discarded Points for which $F^s<0$ have been discarded as well.
Since we set the insertions $\left(\delta^D_{LL}\right)^{32}$ and
$\left(\delta^D_{RR}\right)^{32}$ equal to zero the increase of
$BR(B^0_s\rightarrow\mu^+\mu^-)$ compared to the SM prediction seen in
the left panel of figure \ref{fig:inser} is mainly due to the effects
discussed in section 4.

It is clear form figure \ref{fig:inser} that even with all the cuts
imposed the possible values of the branching ratio
$BR(B^0_d\rightarrow\mu^+\mu^-)$ can still be above the present
experimental limit (\ref{eqn:Bll_lim}) which means that the
non-observation of the $B^0_d\rightarrow\mu^+\mu^-$ decay imposes
nontrivial constraints on the MSSM parameter space and on the mass
insertions $\left(\delta^D_{LL}\right)^{31}$ and
$\left(\delta^D_{RR}\right)^{31}$.

Finally, comparison the possible effects in the MSSM without and with 
flavour violation in the squark mass matrices (figures \ref{fig:corel} 
and \ref{fig:inser}, respectively) leads to the interesting
conclusion that, within the supersymmetric framework, observation of
the $B^0_d\rightarrow\mu^+\mu^-$ decay at the level close to the
present BaBar limit (\ref{eqn:Bll_lim})
(i.e. with BR above $\sim6\times10^{-8}$ in general and
above $\sim3\times10^{-8}$ if unnaturally large and very unlikely values
of the stop mixing parameters $A_t$ are not taken into account), apart 
for implying that the scale of the Higgs sector
is not far from the electroweak scale, would be a very strong evidence of
non-minimal flavour violation in the quark sector.

\section{Effects of flavour violation in the lepton sector}

To complete the picture of flavour violation in the supersymmetric
extension of the SM model we discuss briefly also the lepton sector.

To account for the observed atmospheric and solar neutrino
oscillations \cite{SUPERK} the analog of the CKM mixing matrix, the
so-called Maki-Nakagawa-Sakata mixing matrix \cite{MNS}, has to be
introduced in the leptonic sector of the SM or the MSSM. Under the
assumption that the mixing occurs only between the three know neutrino
flavours (no sterile neutrinos) which is supported by the SNO results
\cite{SNO}, the MNS matrix $U$ is of dimension $3\times3$ and is
usually parameterized in a similar way as the CKM matrix
\begin{eqnarray}
U=\left(\matrix{
c_{12}c_{13} & s_{12}c_{13} & s_{13}\cr
\mbox{\boldmath$\cdot$} &\mbox{\boldmath$\cdot$}& s_{23}c_{13}\cr
\mbox{\boldmath$\cdot$} &\mbox{\boldmath$\cdot$}& c_{23}c_{13}}\right)
\label{eqn:mnsmat}
\end{eqnarray}
where $c_{12}=\cos\theta_{12}$ etc. and where we show only the entries
directly related to observed oscillations and neglected all possible
CP violating phases. Non-zero angles $\theta_{12}$ and $\theta_{23}$
are responsible for solar and atmospheric neutrino oscillations,
respectively.

The pattern of $U$ emerging from the experimental data \cite{NISH,CHOOZ}
\begin{eqnarray}
|U_{11}|\approx |U_{12}|\approx1/\sqrt2
\nonumber\\  
|U_{23}|\approx |U_{33}|\approx1/\sqrt2
\label{eqn:patt}\\  
|U_{13}|\approx0\phantom{aaaa}\nonumber
\end{eqnarray}
is called bi-maximal mixing and is distinctly different from the
hierarchical pattern of the CKM matrix.

Nontrivial mixing matrix $U$ (\ref{eqn:mnsmat}) induces of course also
flavour violating processes with charged leptons, such a
$\mu\rightarrow e\gamma$ or $Z^0\rightarrow e\mu$ etc. but at rates
which are unmeasurably small (e.g. $BR(\mu\rightarrow
e\gamma)<10^{-50}$) for neutrino mass squared differences required to
explain the Superkamiokande data:
\begin{eqnarray}
\Delta m^2_{\rm atm}\equiv 
m^2_{\nu_3}-m^2_{\nu_2}\approx3.2\times10^{-3} {\rm ~eV}^2~,
\nonumber\\
\Delta m^2_{\rm sol}\equiv 
m^2_{\nu_2}-m^2_{\nu_1}\sim{\cal O}\left(10^{-4}\right) {\rm ~eV}^2
\phantom{aa}
\nonumber
\end{eqnarray}
and masses compatible with constraints imposed by cosmology ($\sum_a
m_{\nu_a}<$ few eV). Thus, if the MNS mixing matrix is the only source
of flavour violation in the leptonic sector, neutrino oscillations
remain the only observable lepton flavour violating phenomenon.

In supersymmetric extension of the SM lepton flavour violation can
originate also in the slepton mass squared matrices.  Existing
experimental upper bounds: $BR(\mu\rightarrow e\gamma)<10^{-11}$,
$BR(\tau\rightarrow e(\mu)\gamma)<10^{-6}$ put stringent constraint
only on the mass insertion $|\left(\delta_{LR}^l\right)^{12}|$ which
has to be smaller than $10^{-5}$; constraints on the other sleptonic
mass insertions are of order few$\times10^{-1}$ \cite{MIPORO}.

Interesting links between the lepton flavour violation originating in
the slepton sector and neutrino masses and mixing exist in the see-saw
scenario in which observed small neutrino masses result from exchanges
of right-handed neutrinos $\nu_R$ of masses
$M_{\nu_R}\sim10^{10}-10^{14}$ GeV in the GUT-type framework.
Firstly, the RG running of the parameters of the theory between the
scales $M_{\rm GUT}$ and $M_{\nu_R}$ necessarily induces lepton
flavour violating mass insertions. It turns out that the experimental
limits on $\mu\rightarrow e\gamma$, $\tau\rightarrow e(\mu)\gamma$
decays put interesting constraints on realizations of the see-saw
mechanism in the GUT-type scenarios \cite{LAMASA}.  Secondly, lepton
flavour violating originating in the slepton sector can influence
neutrino masses and mixing via quantum corrections below the scale
$M_{\nu_R}$. Let us discuss this point in some details.

Quantum corrections to neutrino masses and mixing below the
$M_{\nu_R}$ scale are of two types. The first one are the corrections
depending on $\ln(M_{\nu_R}/M_W)$ which are accounted for by
integrating the renormalization group equations \cite{RGE} of the
dimension 5 operator between the scales $M_{\nu_R}$ and $M_W$. The
most interesting aspect of the RG equations is their fixed point
structure \cite{FP}: whenever the RG running is substantial, the
mixing angles evolve in such a way that at the $M_W$ scale either
$U_{31}=0$ or $U_{32}=0$. In both cases one gets the following
relation between the mixing angles
\begin{eqnarray}
\sin^22\theta_{12} = {s^2_{13}\over(s^2_{23}c^2_{13}+s^2_{13})^2}
\sin^22\theta_{23}\nonumber
\end{eqnarray}
Because of the CHOOZ limit $s^2_{13}<0.16$ \cite{CHOOZ} this is
incompatible with the bi-maximal mixing pattern (\ref{eqn:patt})
favoured by the solar and atmospheric neutrino data. This means that
always for exact three-fold or two-fold degeneracy of neutrino masses
at the scale $M_{\nu_R}$, or for approximate degeneracies of neutrinos
having the same CP parities, when the RG running is substantial
\cite{FP}, the mixing angles obtained from the see-saw mechanism are
phenomenologically unacceptable.  Note also that these are precisely
the most interesting cases: three-fold degeneracy of neutrinos will be
required if the neutrinoless double beta decay is found at the level
requiring $m_\nu^{ee}\sim0.5$ eV. More generally, see-saw scenarios
giving naturally large mixing angles may be easier to find if the
spectrum of neutrino masses is (approximately) degenerate.

The unacceptable pattern of mixing generated by RG running can however
be changed by the second type of quantum corrections to the neutrino
mass matrix - the so-called low energy threshold corrections - if
there is some lepton flavour violation in the slepton sector
\cite{CHUPO}.

In the basis in which the neutrino mass matrix 
$(\mbox{\boldmath$m$}_\nu^0)^{AB}$ generated by the underlying see-saw 
mechanism is diagonal the corrected neutrino mass matrix can be written 
as \cite{CHWA,CHPOK}
\begin{eqnarray}
m_{\nu_a}^{(0)}\delta^{ab}
+\left[U^T\left(\mbox{\boldmath$I$}^T\mbox{\boldmath$m$}_\nu^{(0)}
+\mbox{\boldmath$m$}_\nu^{(0)}\mbox{\boldmath$I$}\right)
U\right]^{ab}\label{eqn:mnucorr}
\end{eqnarray}
where $U$ is the uncorrected MNS matrix and
\begin{eqnarray}
\mbox{\boldmath$I$}^{AB} = \mbox{\boldmath$I$}^{AB}_{\rm th}
-\delta^{AB}\mbox{\boldmath$I$}^A_{\rm rg}
\end{eqnarray}
summarizes the RG ($\mbox{\boldmath$I$}^A_{\rm rg}$) and low energy
threshold ($\mbox{\boldmath$I$}^{AB}_{\rm th}$) corrections. The most
interesting part of the latter corrections take the form \cite{CHWA}
\begin{eqnarray}
\mbox{\boldmath$I$}^{AB}_{\rm th}\approx
\left(\delta^l_{LL}\right)^{AB}\times f(M^2_{\tilde l},m^2_C)
\nonumber
\end{eqnarray}
where the function of chargino and slepton masses $f$ is typically of
order few$\times(10^{-4}-10^{-3})$ (contributions of $\delta^l_{RR}$
and $\delta^l_{LR}$ to $\mbox{\boldmath$I$}^{AB}_{\rm th}$ are
smaller).  For comparison, for $M_{\nu_R}\sim10^{10}$ GeV,
$\mbox{\boldmath$I$}^\tau_{\rm rg}\approx10^{-5}\times\tan^2\beta$,
and $\mbox{\boldmath$I$}^\mu_{\rm rg}$, $\mbox{\boldmath$I$}^e_{\rm
  rg}$ are negligible.

In the case of the (approximate) degeneracy of the zeroth order
neutrino masses $m^{(0)}_{\nu_a}\approx m^{(0)}_{\nu_b}$ the matrix
$U$ is fixed by the condition
\begin{eqnarray}
\sum_{AB}U_{Aa}\mbox{\boldmath$I$}_{AB}U_{Bb}=0\label{eqn:fplike}
\end{eqnarray}
(the freedom $U\rightarrow U\cdot R_{ab}$, where $R_{ab}$ is an
arbitrary rotation of the $\nu_a$ and $\nu_b$ neutrino fields, is used
to diagonalize the ``perturbation'').  This leads to the fixed
point-like relations between the mixing angles which are different
than the RG evolution provided $|\mbox{\boldmath$I$}^{AB}_{\rm
  th}|\simgt|\mbox{\boldmath$I$}^A_{\rm rge}|$ \cite{CHUPO}.

As an example consider initial degeneracy of the three neutrinos
$m_{\nu_a}\approx m_{\nu_b}\approx -m_{\nu_c}$ and only one dominant
correction $\mbox{\boldmath$I$}^{AB}_{\rm th}$. In this case
interesting results are obtained for $m_{\nu_1}\approx m_{\nu_3}$ (or
$m_{\nu_2}\approx m_{\nu_3}$) and dominant
$\mbox{\boldmath$I$}^{\mu\tau}_{\rm th}$ correction (i.e.
$(\delta^l_{LL})^{23}\neq0$). The condition (\ref{eqn:fplike}) then
gives
\begin{eqnarray}
s_{13}=-\cot2\theta_{23}\tan\theta_{12} ~(\cot\theta_{12})
\nonumber
\end{eqnarray}
which is compatible with the bi-maximal mixing and small $U_{13}=s_{13}$ 
element. Moreover, for the mass squared differences one obtains
\begin{eqnarray}
&&\Delta m^2_{\rm sol}=-4m^2_\nu\cos2\theta_{12}\sin2\theta_{23}
\mbox{\boldmath$I$}^{\mu\tau}_{\rm th}\nonumber\\
&&\Delta m^2_{\rm atm}=-4m^2_\nu(1+\cos^2\theta_{12})\sin2\theta_{23}
\mbox{\boldmath$I$}^{\mu\tau}_{\rm th}\nonumber
\end{eqnarray}
that is, for the bi-maximal mixing:
\begin{eqnarray}
\Delta m^2_{32}\gg\Delta m^2_{21}\sim0
\nonumber
\end{eqnarray}
in agreement with the experimental information.  $\Delta m^2_{\rm
  atm}\approx3\times10^{-3}$ eV$^2$ requires then $m_\nu\approx1$
eV$^2$ and $(\delta^l_{LL})^{23}\sim0.5$ with the interesting
implications for the $\tau\rightarrow\mu\gamma$ decay. $\Delta
m^2_{\rm sol}$ of right magnitude can be generated either by departure
from $\theta_{12}=\pi/4$ or, by another, hierarchically smaller,
correction: flavour conserving
$\mbox{\boldmath$I$}^{AB}=\mbox{\boldmath$I$}^A\delta^{AB}$ with
either $\mbox{\boldmath$I$}^\tau\neq0$ or
$\mbox{\boldmath$I$}^\mu\neq0$ (e.g. $\mbox{\boldmath$I$}^\tau\neq0$
from RG running for not too large value of $\tan\beta$), or flavour
violating correction $\mbox{\boldmath$I$}^{e\mu}_{\rm th}$, or
$\mbox{\boldmath$I$}^{e\tau}_{\rm th}$.

There can be, of course, other interesting cases with more complicated
interplay of RG and low energy threshold corrections \cite{CHU,CHPOK}.

\section{Summary}

We have reviewed recent developments in exploring flavour dynamics in
the supersymmetric extension of the Standard Model. Emphasis has been
put on possible interesting effects in $b$-physics arising for large
values of $\tan\beta$ and not too high a scale of the MSSM Higgs boson
sector, both in the case of minimal flavour violation and in the case
of flavour violation originating in the sfermion sector. We have
discussed the importance of the flavour changing neutral Higgs boson
couplings generated by the scalar penguin diagrams and their role in
constraining the MSSM parameter space.  We have shown that in the case
of minimal flavour violation the experimental lower limit on 
$B^0_s$-$\bar B^0_s$
mass difference constrains branching fractions of the decays
$B^0_{d,s}\rightarrow\mu^+\mu^-$ possible in the MSSM. We have also
pointed out that observation of the $B^0_d\rightarrow\mu^+\mu^-$ decay
with BR at the level $\simgt3\times10^{-8}$ 
(and even lower if $\Delta M_s$ turns out to be bigger than 15/ps)
would be a strong indication of
nonminimal flavour violation in the quark sector. Flavour violation
connected with neutrino oscillations has been also discussed. It has
been argued that in some physically interesting situations flavour
violation originating in the slepton mass matrices can be responsible
(at least in part) for observed pattern of the neutrino mixing and
mass squared differences.

\vskip 0.5cm

\noindent {\bf Acknowledgments}
\vskip 0.3cm
\noindent We would like to thank A.J. Buras and \L . S\l awianowska
in collaboration with whom some of the results presented here have
been worked out. The work was partly supported by the Polish State
Committee for Scientific Research grant 5 P03B 119 20 for 2001-2002
and by the EC Contract HPRN-CT-2000-00148 for years 2000-2004.  The
work of J.R. was also supported by the German Bundesministerium f\"ur
Bildung und Forschung under the contract 05HT1WOA3 and the Deutsche
Forschung Gemainschaft Project Bu. 706/1-1.  

\end{document}